\newcommand{\NPA}[3]{Nucl.\ Phys.\ A\ {\bf #1},\ #2 (#3)}

\newcommand{\PLB}[3]{Phys.\ Lett.\ B\ {\bf #1},\ #2 (#3)}

\newcommand{\PRL}[3]{Phys.\ Rev.\ Lett.\ {\bf #1},\ #2 (#3)}

\newcommand{\PRC}[3]{Phys.\ Rev.\ C\ {\bf #1},\ #2 (#3)}
\newcommand{\PRD}[3]{Phys.\ Rev.\ D\ {\bf #1},\ #2 (#3)}

\newcommand{\diracslash}[1]{#1\llap{/\kern2pt}}

\newcommand{\be}{\begin{equation}}
\newcommand{\ee}{\end{equation}}
\newcommand{\bea}{\begin{eqnarray}}
\newcommand{\eea}{\end{eqnarray}}
\newcommand{\ba}[1]{\begin{array}{#1}}
\newcommand{\ea}{\end{array}}

\documentclass[prd,aps,floats,nofootinbib,tightenlines,showpacs]{revtex4}
\usepackage{epsfig,graphicx,pstricks}
\usepackage{psfrag}
\usepackage{color}
\usepackage{amsmath}
\usepackage{amsfonts}
\usepackage{amssymb}
\usepackage{textcomp}
\usepackage{multirow}
\usepackage{subfigure}
\usepackage{slashed}

\begin{document}

\title {Curing the acausal behavior of the sound velocity in an excluded volume hadron resonance gas model}
\author{Guru Prakash Kadam }
\email{guruprasad@prl.res.in}
\affiliation{Theory Division, Physical Research Laboratory,
Navrangpura, Ahmedabad 380 009, India}

\date{\today} 

\def\be{\begin{equation}}
\def\ee{\end{equation}}
\def\bearr{\begin{eqnarray}}
\def\eearr{\end{eqnarray}}
\def\zbf#1{{\bf {#1}}}
\def\bfm#1{\mbox{\boldmath $#1$}}
\def\hf{\frac{1}{2}}
\def\sl{\hspace{-0.15cm}/}
\def\omit#1{_{\!\rlap{$\scriptscriptstyle \backslash$}
{\scriptscriptstyle #1}}}
\def\vec#1{\mathchoice
        {\mbox{\boldmath $#1$}}
        {\mbox{\boldmath $#1$}}
        {\mbox{\boldmath $\scriptstyle #1$}}
        {\mbox{\boldmath $\scriptscriptstyle #1$}}
}

\begin{abstract}
We improve the excluded volume hadron resonances gas model (EHRG) to cure the acausal behavior of the sound velocity which is typical of excluded volume models. We achieve this by including temperature (T) and density ($\mu$) dependent hadron masses in the partition function of EHRG. The temperature and density dependent masses of the constituent quarks (u,d,s) and the light mesons are obtained within Nambu-Jona-Lasinio model while for the heavy hadrons we use linear scaling rule in terms of constituent quarks.  With this improvement, we observe that the velocity of sound flattens at high temperature unlike old EHRG models where the sound velocity rises very rapidly at high temperatures which is the indication of its super-luminal behavior. 
\end{abstract}


\maketitle

\section{Introduction}
The hadron resonance gas model (HRG) is based on the premise that the interacting gas of hadrons and resonances can be approximated by that of ideal gas of hadrons and resonances provided all the resonance states are included in the partition function\cite{Huovinen} and the resonance width are small enough so that they can be treated as an elementary particles. In the S-matrix formulation of statistical mechanics this is an exact theorem known as Dashen-Bernstein-Ma theorem\cite{DBM} and it is the legitimate way to account for the attractive interactions. This is true in the simplest situation, as shown if Ref.\cite{welke,prakashvenu}, where the authors showed that the thermodynamical quantities for the gas of interacting pions calculated using relativistic virial expansion with an experimental phase shift nearly coincide with that of ideal gas of pions and $\rho$ resonances. But this formulation certainly misses repulsive interactions between hadrons, especially baryons,  which are important too while estimating the thermodynamical quantities of hadronic matter. Such repulsive interaction can be accounted in to the model by Van-der-Waals excluded volume approach in the molecular physics.  In this formulation the volume in the partition function is substituted by the effective volume obtained by subtracting the volume occupied by the particles in the system due to their finite size. There are two thermodynamically consistent excluded volume formulation of HRG exist in the literature, $viz.$, HRG with constant excluded volume\cite{ehrgrishke,ehrgclaymans,yen,gorengreinerjpg,gorengreinerprc,gorensteinplb,gorensteinjpg,yengorenprc} and HRG with excluded volume in the MIT bag picture\cite{kapustaolive}. Such excluded volume extensions of HRG has been successful in explaining total particle number density at freeze-out\cite{yen}. Further, these models has been confronted with the lattice QCD\cite{andronic,kapusta,vovchenko} and has been used as practical model to estimate the transport properties of hadronic matter\cite{gurunpa,gurumpla,guruprc,greinerprc,greinerprl}.

Despite its success, excluded volume HRG models are plagued by the acausal behavior of the sound velocity\cite{prakashvenu}. This acausal behavior is typical of all excluded volume models which is the indicator of the first order liquid-gas phase transition in the model. When the gas approaches the incompressible liquid phase, any disturbance in the system propagate through the medium almost instantaneously, thus violating the principle of causality. Thus, the excluded volume models break down at high temperature and especially at high densities.  One possible remedy is to account for the relativistic contraction of the hard spheres in the excluded volume models  which is rather technically complicated\cite{Bugaev}. Recently, we proposed an improvement in an excluded volume HRG (EHRG) by accounting the temperature (T) and density ($\mu$) dependence  of the hadrons\cite{guruarxive}. This improvement is technically simple but physically more appealing.  We argued that before computing any thermodynamical quantity using HRG model or any of its extension, one needs to account for T and $\mu$ dependence of hadrons which is the upshot of the chiral symmetry of quantum chromodynamics (QCD); symmetry which should be respected by any of QCD's effective model. We observed that accounting such T and $\mu$ dependence of the hadron masses, the thermodynamics of HRG model is non-trivially altered especially at moderately high temperatures. We further observed the rapid rise of interaction measure as well as entropy density in the improved EHRG  as compared to normal EHRG model. Since sound velocity is a derived quantity and as it is intimately related to pressure, energy density and entropy density, non trivial change observed in these quantities would certainly affect the behavior of the sound velocity.

In this work we use improved EHRG model to see the behavior of sound velocity, especially its causal structure. We see that including T and $\mu$ dependent hadron masses in EHRG model, sound velocity remains causal at high temperatures. To compute T and $\mu$ dependent hadron masses we use linear scaling rule in terms of constituent quarks. The masses of the constituent quarks (u,d,s) and that of light mesons are computed using Nambu-Jona-Lasinio model at finite T and $\mu$.

We organize the paper as follows. In Sec. II we briefly describe the Nambu-Jona-Lasinio model at finite temperature and baryon density. In Sec. III we describe our improvement in the old excluded volume hadron resonance gas (EHRG model. In Sec. IV we discuss the thermodynamics of hadron gas within ambit of improved EHRG model. Finally we summarize and conclude in Sec. V.

\section{SU(3) Nambu-Jona-Lasinio model (NJL) at finite temperature and baryon density}
In this section we just recapitulate the NJL model formalism described in Ref.\cite{costa} keeping all the notations but the constituent quark mass which we denote by $\mathcal{M}$ while we reserve the notation  M for the mass of hadrons. For more details, see Refs.\cite{klevansky,hatsuda,vogl}. The NJL model with three quark flavors (u,d,s) is described by the Lagrangian density,
 \be
 \mathcal{L}=\bar q(i\slashed \partial-\tilde m)q+\frac{g_{S}}{2}\sum_{a=0}^{8}\bigg[(\bar q\lambda^{a}q)^{2}+(\bar qi\gamma\lambda^{a}q)^{2}\bigg]+g_{D}\bigg\{\text{det}[\bar q(1+\gamma_{5})q]+\text{det}[\bar q(1-\gamma_{5})q]\bigg\}
 \label{njl_lagrangian}
 \ee
 Here q=(u,d,s) is the quark triplet, $\tilde m=(m_{u},m_{d},m_{s})$ is the current quark mass matrix which explicitly breaks SU(3) chiral symmetry explicitly. $\lambda^{a}$'s are Gell-Mann matrices. The determinant term is chiral invariant but breaks U$_A$(1) symmetry and is the reflection of the axial anomaly in QCD.
 
 Bosonization of Lagragian (\ref{njl_lagrangian}) leads to an effective action given by
  \be
  W_{eff}[\varphi,\sigma]=-\frac{1}{2}(\sigma^{a}\text{S}^{-1}_{ab}\sigma^{b})-\frac{1}{2}(\varphi^{a}\text{P}^{-1}_{ab}\varphi^{b})-i\: \text{Tr ln} [i(\gamma_{\mu}\partial_{\mu})-\tilde m+\sigma_{a}\lambda^{a}+(i\gamma_{5})(\varphi_{a}\lambda^{a})]
  \label{eff_act}
  \ee
  Here, $\sigma^{a}$ and $\varphi^{a}$ are the scalar and pseudo-scalar fields respectively.  S$_{ab}$ and P$_{ab}$ are the projection operators defined as
\begin{eqnarray}
\text{S}_{ab} &=& g_S \delta_{ab} + g_D \text{D}_{abc}<\bar{q} \lambda^c q>, \label{sab}\\
\text{P}_{ab} &=& g_S \delta_{ab} - g_D \text{D}_{abc}<\bar{q} \lambda^c q>. \label{pab}
\end{eqnarray}
 
 D$_{abc}$ are the SU(3) structure constants for $a,b,c=1,2,...8$ while, D$_{0ab}=-(1/\sqrt{6})\delta_{ab}$ and D$_{000}=\sqrt{2/3}$. 
 
 Variation of the action (\ref{eff_act}) leads to the constituent quark masses as,
 \be
 \mathcal{M}_i = m_i - 2g_{_S} <\bar{q_i}q_i> -2g_{_D}<\bar{q_j}q_j><\bar{q_k}q_k>
\ee
 with the cyclic permutation of $i,j,k=u,d,s$.
 One can expand the effective action (\ref{eff_act}) over mesonic fields 
 \be
W_{eff}^{(2)}[\varphi] =
-\frac{1}{2}\varphi^a \left[ P_{ab}^{-1} - \Pi_{ab} (P) \right] \varphi^b
= -\frac{1}{2}\varphi^a  D_{ab}^{-1}(P)  \varphi^b
\ee
Here, $D_{ab}$ is the non-normalized meson propagator. Here we have kept pseudo-scalar mesons only.  The polarization operator $\Pi_{ab}$ is defined as
 
\begin{eqnarray}\label{polop}
\Pi_{ab} (P) = i N_c \int \frac{d^4p}{(2\pi)^4}\mbox{tr}_{D}\left[
S_i (p) (\lambda^a)_{ij} (i \gamma_5 )
S_j (p+P)(\lambda^b)_{ji} (i \gamma_5 )
\right],
\end{eqnarray}

  From the pole structure of $[1-P_{ij}\Pi^{ij}(P_{0}=\mathcal{M},\vec{P}=0)]=0$, one can obtain the meson masses. For non-diagonal mesons (pions, kaons) the polarization operator can be written as
 \begin{equation}
	\Pi^{ij} (P_0) = 4 \{(I_1^i + I_1^j)-[P_0^2-(\mathcal{M}_i-\mathcal{M}_j)^2]\,\,I_2^{ij}(P_0)\}
\end{equation}
Here, the integrals $I_{1}^{i}$ and $I_{2}^{i,j}$ at $T=0$ and $\mu=0$ are defined as,
 \begin{eqnarray}
I_1^i &=& i N_c \int \frac{d^4p}{(2\pi)^4} \, \frac{1}{p^2-\mathcal{M}_i^2}
       = \frac{N_c}{4 \pi^2} \int^{\Lambda}_0 \frac{{\tt p}^2 d {\tt p}}{E_i} ,
\end{eqnarray}

\begin{eqnarray}
I_2^{ij}(P_0) &=& i N_c \int \frac{d^4p}{(2\pi)^4} \, \frac{1}{(p^2-\mathcal{M}_i^2)((p+P_0)^2-\mathcal{M}_j^2)}
\nonumber \\
       &=& \frac{N_c}{4 \pi^2} \int^{\Lambda}_0 \frac{{\tt p}^2 d {\tt p}}{E_i E_j}
       \,\,\, \frac{E_i+E_j}{P_0^2-(E_i+ E_j)^2} \, ,
\end{eqnarray}
where $E_{i,j}=\sqrt{{\tt p}^2+\mathcal{M}_{i,j}^2}$  and $\Lambda$ is the O(3) cut-off. At $T\neq0$ and $\mu\neq0$, these integrals are non-trivially modified and involve Fermi-Dirac distribution function, $f^{\pm}=\frac{1}{1+e^{(E\pm\mu)/T}}$.
 To compute $\eta$ and $\eta^{'}$ masses, we consider the matrix representation of the operators $P_{ab}$ and $\Pi_{ab}$ in the basis of $\pi^{0}$-$\eta$-$\eta^{'}$ system,
 \begin{eqnarray}\label{Pab}
	&&  {P}_{ab} =
  	\left(
		\begin{array}{ccc}
			P_{33} & P_{30} & P_{38} \\
			P_{03} & P_{00}& P_{08} \\
			P_{83} & P_{80}& P_{88}
		\end{array}
	\right)\,\,\,\,\,\,\mbox{and}\,\,\,\,\,\,{\Pi}_{ab} =
  \left(
	\begin{array}{ccc}
		\Pi_{33} & \Pi_{30} & \Pi_{38} \\
		\Pi_{03} & \Pi_{00}& \Pi_{08} \\
		\Pi_{83} & \Pi_{80}& \Pi_{88}
	\end{array}
  \right).
\end{eqnarray}
$\eta$ and $\eta^{'}$ masses can obtained from the inverse propagators,
\begin{eqnarray}
	D_\eta^{-1} (P) =\left( { A}+{ C}\right) -  \sqrt{({ C}-{ A})^2+4{ B}^2} 
 	\\
	D_{\eta'}^{-1}(P) =\left( { A}+{ C}\right) +  \sqrt{({ C}-{ A})^2+4{ B}^2} 
\end{eqnarray}
with ${ A} = P_{88} -\Delta \Pi_{00}(P),
      { C} = P_{00} -\Delta \Pi_{88}(P),
      { B} = - (P_{08} +\Delta \Pi_{08}(P))$ and
     $\Delta = P_{00} P_{88}- P_{08}^2 $. 
     
 The matrix elements $P_{33}$, $P_{00}$, $P_{88}$ and $P_{08}$ can be expressed in terms of quark condensates ($\langle\bar q_{i}q_{i}\rangle$),
     \begin{eqnarray}
P_{33}&=& g_S +             g_D <\bar{q}_{s}\,q_s>   , \\
P_{00}&=&g_{S}-\frac{2}{3}g_{D}\left( <\bar{q}_{u}\,q_{u}>+<\bar{q}_{d}\,q_{d}>+<\bar{q}_{s}\,q_{s}>\right) , \\
P_{88}&=&g_{S}+\frac{1}{3}g_{D}\left( 2<\bar{q}_{u}\,q_{u}>+2<\bar{q}_{d}\,q_{d}>-<\bar{q}_{s}\,q_{s}>\right) ,\\
P_{08}&=&P_{80}=\frac{1}{3\sqrt{2}}g_{D}\left( <\bar{q}_{u}\,q_{u}>+<\bar{q}_{d}\,q_{d}>-2<\bar{q}_{s}\,q_{s}>\right),
\end{eqnarray}
Again at finite temperature and baryon density these expressions involve integrals over Fermi-Dirac distribution function\cite{klevansky,hatsuda}.

\section{Improved excluded volume hadron resonance gas model}
The fundamental quantity of any statistical model is the partition function. In the hadron resonance gas model, the partition function is just that of an ideal gas summed over all the hadronic species in the system. For the gas of  hadrons contained in volume V at temperature T and baryon chemical potential $\mu$, the partition function reads 
\be
\log\mathcal{Z}(T,\mu,V)=\int dm(\rho_{M}(m)\log Z_{M}(m,T,\mu)+\rho_{B}(m)\log Z_{B}(m,T,\mu))
\label{pfhrg}
\ee
where, $Z_{M}$ and $Z_{B}$ are the partition functions of mesons and baryons respectively. The hadron properties enter the model through the spectral densities $\rho_{M,B}$. Since the gas is assumed to be non interacting, the spectral densities are just Dirac delta functions.
\be
\rho_{M,B}=\sum_{a}^{M_{a}<\Lambda}g_{a}\delta(m-M_{a})
\label{massspectrum}
\ee
Once we know the partition function [Eq. (\ref{pfhrg})] together with the spectral density [Eq. (\ref{massspectrum})], all the thermodynamical quantities can be readily obtained, $viz.$, Pressure $P(T,\mu)=T \lim_{V\rightarrow \infty}$ $\log\mathcal{Z}(T,\mu)/V$, Baryon number density $n_{B}=\partial P(T,\mu)/\partial \mu$, entropy density $s(T,\mu)=\partial P(T,\mu)/\partial T$, energy density $\varepsilon(T,\mu)=Ts(T,\mu)-P(T,\mu)+\mu n_{B}(T,\mu)$.

Another important thermodynamic quantity which is the characteristic of the equation of state is the sound velocity C$_s^{2}$. This quantity governs the the propagation of the small disturbances through the medium and hence strongly depends on the interactions in the system as well as external parameters like temperature and density. Sound velocity can be defined as
\be
C_{s}^{2}=\frac{\partial P}{\partial \epsilon}\bigg\rvert_{\mu}
\ee
 
The hadron resonance gas  model can further be extended to include the repulsive interactions through the excluded volume corrections. In this section we just describe constant excluded volume HRG.  In the thermodynamically consistent constant excluded volume approach, the system volume V is substituted by V$-v$N in the partition function. Here $v$ is the proper volume parameter for the particle with the hard core radius $r_h$. Thus, the volume available for the hadrons is just the system volume less the total volume occupied by all the hadrons due to their finite spacial extension. With this substitution one can obtains the transcendental equation for pressure as\cite{ehrgrishke}
\be
P^{EV}(T,\mu)=\sum_{a}P^{id}(T,\tilde\mu)
\ee
where $P^{id}$ is the ideal gas pressure and $\tilde\mu=\mu-vP^{EV}(T,\mu)$ is the effective chemical potential. Once the pressure is known all the thermodynamical quantities can be readily obtained. The number density, energy density and entropy density are 

\be
n^{EV}(T,\mu)=\sum_{a}\frac{n^{id}_{a}(T,\tilde\mu)}{1+\sum_{a}v_{a}n_{a}^{id}(T,\tilde\mu)}
\ee
\be
\epsilon^{EV}(T,\mu)=\sum_{a}\frac{\epsilon^{id}_{a}(T,\tilde\mu)}{1+\sum_{a}v_{a}n_{a}^{id}(T,\tilde\mu)}
\ee
\be
s^{EV}(T,\mu)=\sum_{a}\frac{s^{id}_{a}(T,\tilde\mu)}{1+\sum_{a}v_{a}n_{a}^{id}(T,\tilde\mu)}
\ee

The quantity ($\Gamma^{-1}={1+\sum_{a}v_{a}n_{a}^{id}(T,\tilde\mu)}$) is the suppression factor typical of any excluded volume model and is always less than one. Thus, any thermodynamical quantity computed within EHRG model is always less than that of non interacting HRG model. For the temperature range in which we are interested, the Boltzmann approximation is rather a good approximation. In this approximation, excluded volume correction merely correspond to an additional factor of exp$(-vP^{EV}/T)$ multiplying the ideal gas pressure.

We improve EHRG model further by including temperature and baryon density dependent hadron masses in the partition function. Since it is rather difficult to obtain T and $\mu$ dependent masses of all the hadrons and their resonances except for light mesons, we use the linear scaling rule for mesons and baryons in terms of their constituent quarks\cite{leupold,jankowski}. Since hadrons are made of either two or three quarks, we write the scaling rule for hadron masses as
\be
 M_{h}(T,\mu)=(N_{q}-N_{s})\mathcal{M}_{q}(T,\mu)+N_{s}\mathcal{M}_{s}(T,\mu)+\kappa_{h}
 \label{scalingrule2}
 \ee
 Here, $\mathcal{M}$ is the constituent quark mass, $N_{q}$ is the number of light quarks in a given hadron and $N_{s}$ is the measure of strangeness content of the hadron. $\kappa_{h}$ is the constant depends on the state but not on the current quark masses.
 
We further separate zero temperature and zero density part $M(T=0$, $\mu=0)$  in Eq. (\ref{scalingrule2}) and absorb $\kappa_{h}$ in it to get
\be
 M_{h}(T,\mu)= M_{h}(T=0,\mu=0)+(N_{q}-N_{s})\mathcal{M}_{q}^{'}(T,\mu)+N_{s}\mathcal{M}_{s}^{'}(T,\mu)
 \label{scalingrule1}
 \ee
 where $\mathcal{M}_{q,s}^{'}$ is only medium (T and $\mu$) dependent part of the constituent quark mass. The scaling rule given by Eq.(\ref{scalingrule1}) is used for all the hadrons but light mesons. The T and $\mu$ dependence of light mesons as well as that of constituent quarks (u,d,s) can obtained in Nambu-Jona-Lasinio (NJL) formalism at finite temperature and density as described in previous section.

\section{Results and discussion}
 We take all the hadrons and their resonances with the mass cut-off  $\Lambda=2$ GeV in Eq. (\ref{massspectrum}) to compute the thermodynamical quantities in HRG model. To be more specific, we take all the hadrons with the mass cutoff $\Lambda=2.252$ GeV for baryons and 2.011 for mesons\cite{amseler}. For the parametrization of NJL model see Ref. (\cite{klevansky,hufner}). The only remaining unknown parameter in our model is the proper volume parameter $v$ or the hadron hardcore radius $r_h$ for that matter. In the constant volume EHRG model, the hardcore radius is fixed and hence $v=4\frac{4}{3}\pi r_{h}^{3}$. It is customary to use this scheme of parametrization since the value of hard core radius for nucleons can be extracted from the nucleon-nucleon scattering experiment and the same hardcore radius can then be used for all other baryons\cite{heppe}. Since there is no detailed knowledge about short range interaction between mesons, one can still use same hard core radius for mesons as that of baryons since the meson charge radius is the same as that of baryons. But for our purpose we use parametrization scheme where the proper volume parameter $v$ is proportional to the mass of individual hadron\cite{kapusta}. Thus we choose parametrization scheme $v=M_{h}(T,\mu)/\varepsilon_{0}$, where  $\varepsilon_{0}=2$ GeVfm$^{-3}$.  Note that in our model, the proper volume parameter $v$ depends on T and $\mu$.

\begin{figure}[h]
\vspace{-0.4cm}
\begin{center}
\begin{tabular}{c c}
 \includegraphics[width=8cm,height=8cm]{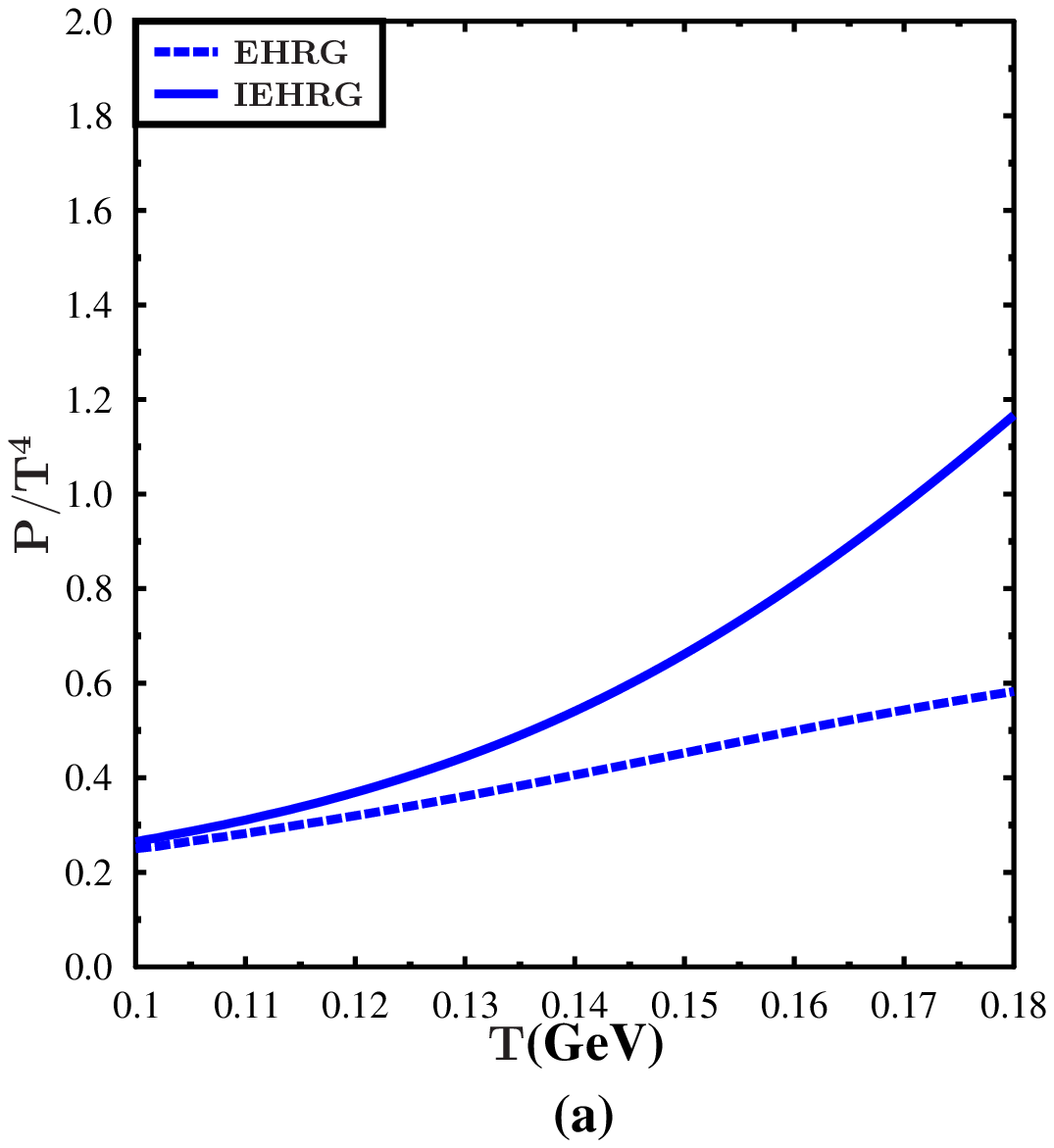}&
  \includegraphics[width=8cm,height=8cm]{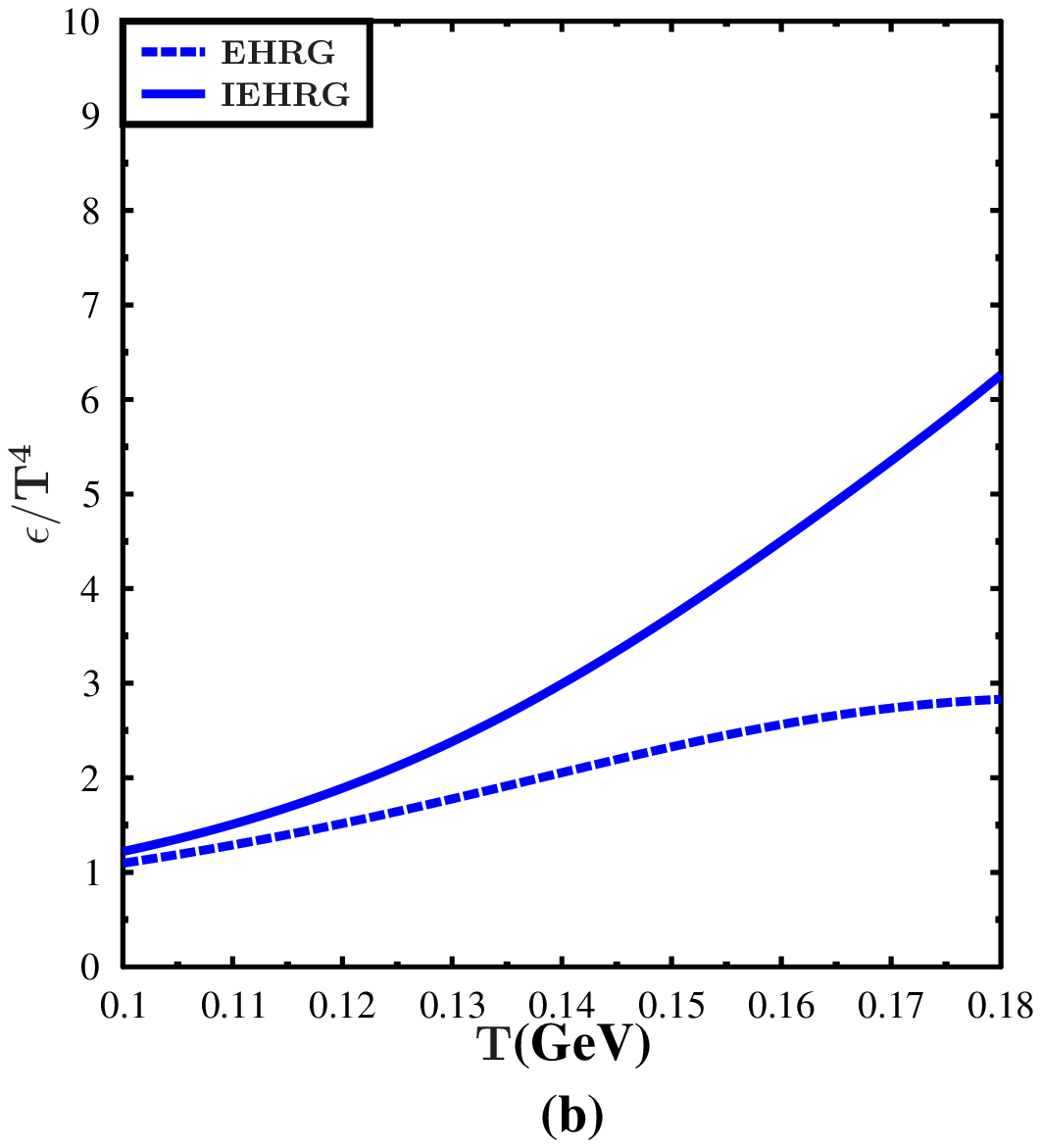}\\
  \end{tabular}
  \caption{(Color online)Results for pressure (left panel) and energy density (right panel) at $\mu=0$ GeV in EHRG (dashed curve) and IEHRG model (solid curve).} 
\label{thermo1}
  \end{center}
 \end{figure}
 
   \begin{figure}[h]
\vspace{-0.4cm}
\begin{center}
 \includegraphics[width=8cm,height=8cm]{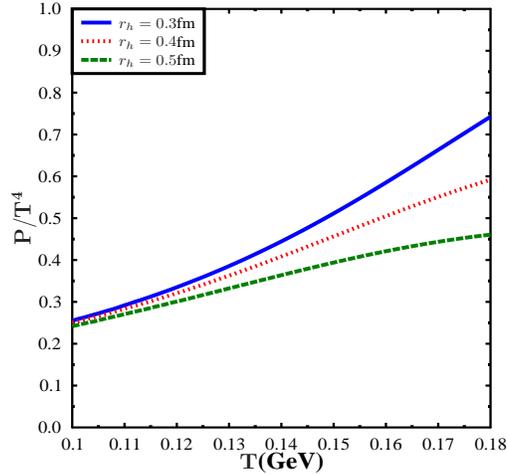}
  \caption{(Color online)Scaled pressure in EHRG model for different values of hadron hardcore radius.} 
\label{thermo2}
  \end{center}
 \end{figure}
 
 We call excluded volume HRG with T and $\mu$ dependent hadron masses as improved excluded volume HRG (IEHRG). [Fig. \ref{thermo1}(a)] shows the pressure (normalized to T$^4$) in EHRG and IEHRG model. We note that the pressure rise more rapidly in IEHRG than EHRG. As mentioned earlier, all the thermodynamical quantities are smaller in excluded volume HRG models than that in HRG model due to the suppression factor ($\Gamma^{-1}$). Also in the constant volume EHRG model for sufficiently high value of the hardcore radius, all the thermodynamical quantities are strongly suppressed at high temperature (see Fig. (\ref{thermo2})). The excluded volume approach was originally devised to suppress the artificially large total particle number densities obtained in HRG model at the chemical freeze-out\cite{yen}. Despite its success in reproducing the correct results for total particle densities at chemical freeze-out, large suppression of the thermodynamical quantities at high temperature become serious flaw of this model. We note from Fig. (\ref{thermo2})  that scaled pressure decreases above T$=$0.150 GeV. As has been noted in Ref.\cite{prakashvenu} this behavior is the first indicative of acausal behavior in excluded volume models. [Fig. \ref{thermo1}(b)] shows the scaled energy density in EHRG and IEHRG model. Again we note that the energy density rises more rapidly in IEHRG than in EHRG.
 
\begin{figure}[h]
\vspace{-0.4cm}
\begin{center}
\begin{tabular}{c c}
 \includegraphics[width=8cm,height=8cm]{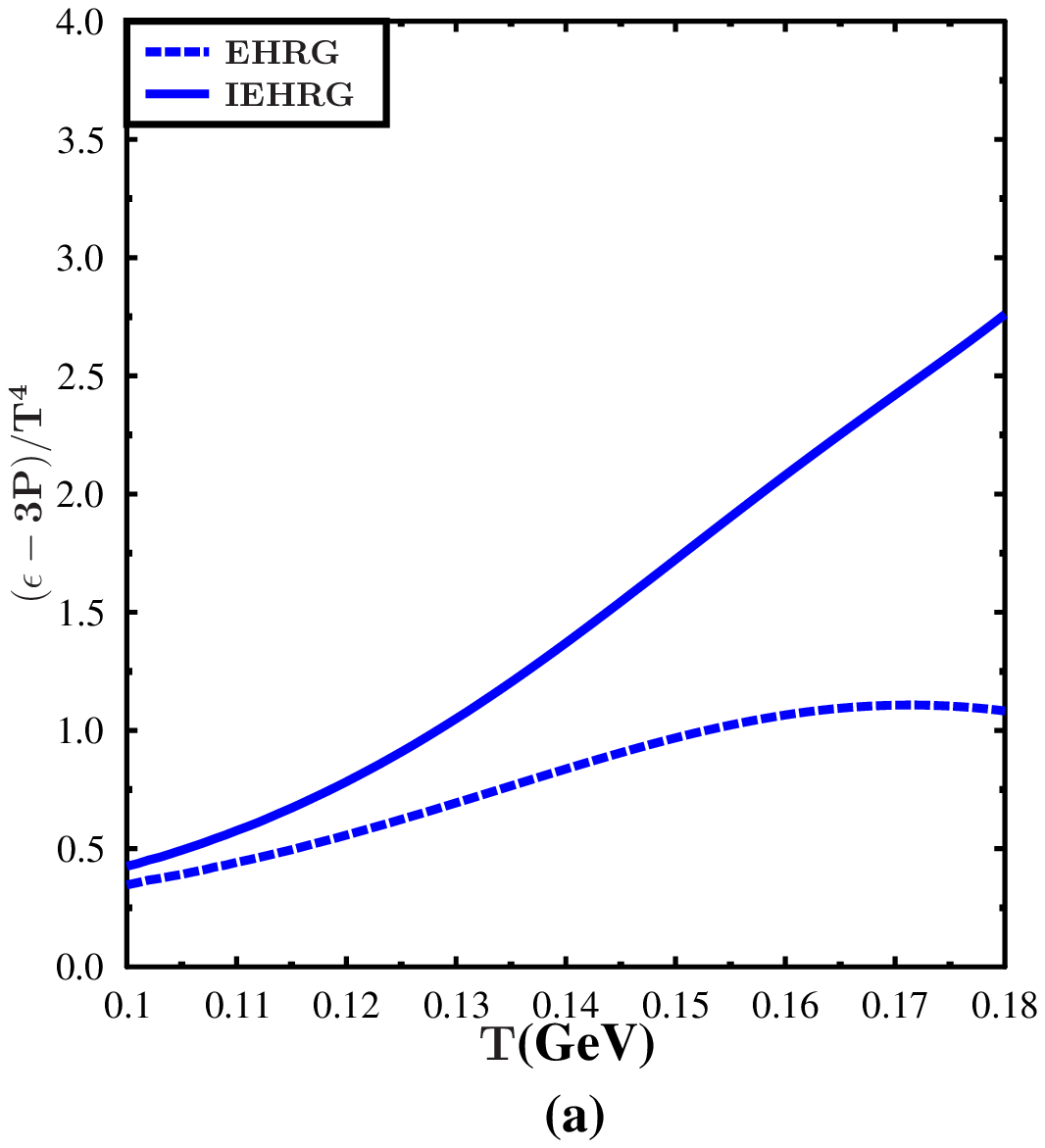}&
  \includegraphics[width=8cm,height=8cm]{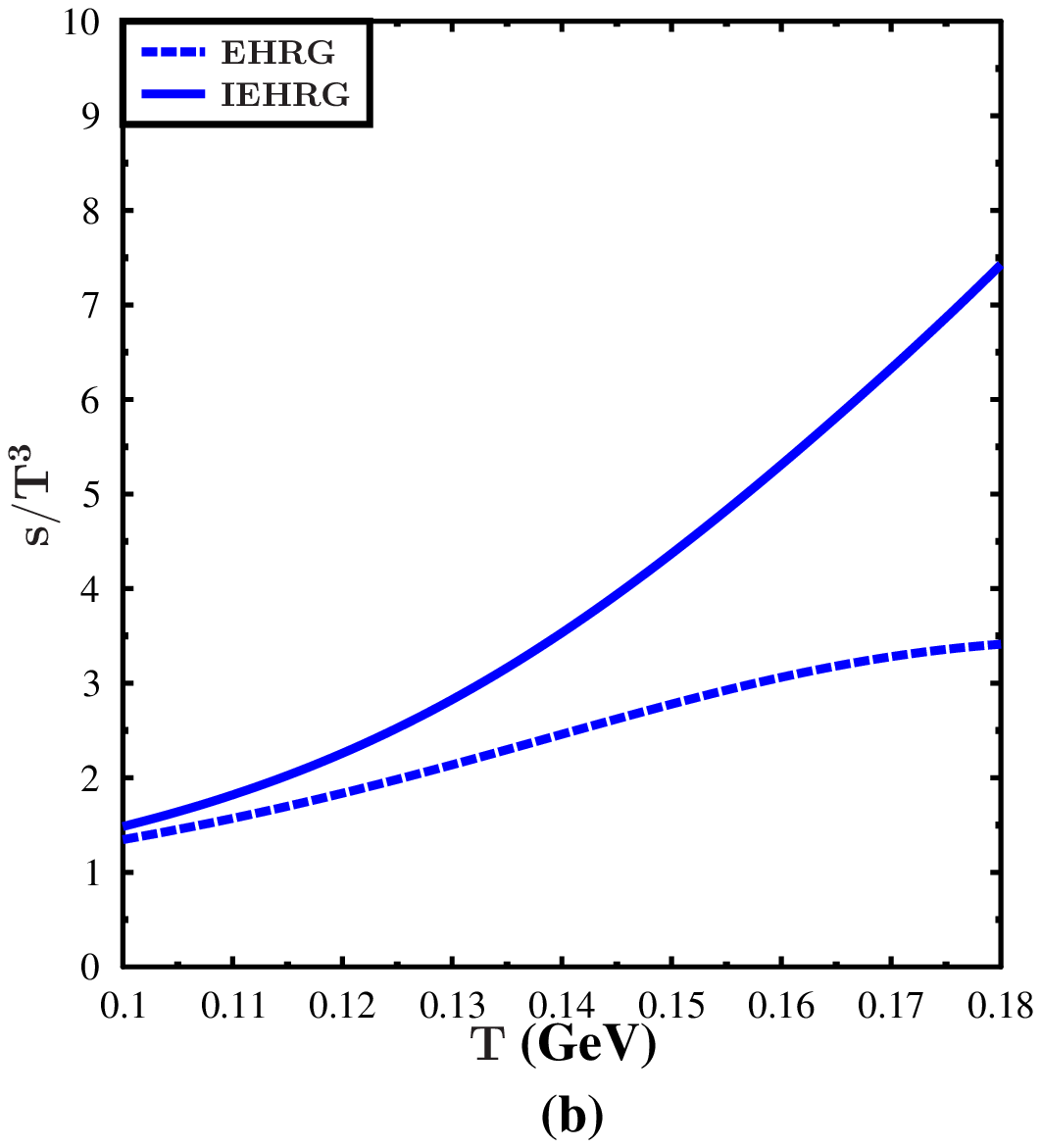}\\
  \end{tabular}
  \caption{(Color online)Results for interaction measure (left panel) and entropy density (right panel) at $\mu=0$ GeV in EHRG (dashed curve) and IEHRG model (solid curve).} 
\label{thermo3}
  \end{center}
 \end{figure}

\begin{figure}[h]
\vspace{-0.4cm}
\begin{center}
 \includegraphics[width=8cm,height=8cm]{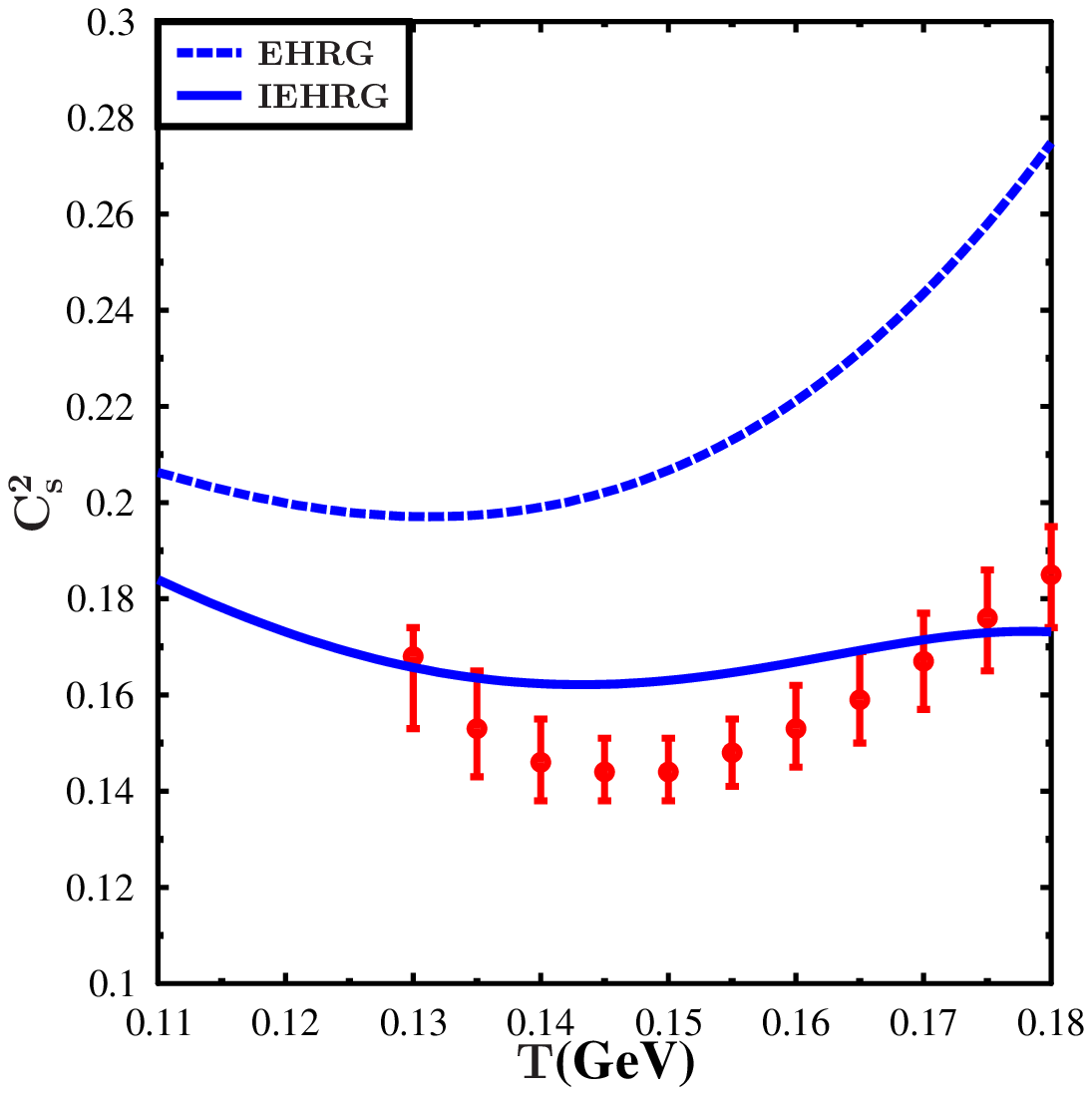}
  \caption{(Color online)Sound velocity in EHRG (dashed curve) and IEHRG model (solid curve) at $\mu=0$ GeV. Lattice data is from\cite{prasad}.} 
\label{thermo5}
  \end{center}
 \end{figure}

 [Fig. \ref{thermo3}(a)] shows the scaled interaction measure (trace anomaly) in EHRG and IEHRG model. We note that while in EHRG interaction measure decreases at high temperature, in IEHRG it rises very rapidly at high temperatures.  Such rapid rise has also been observed in HRG model extended to account for repulsive interaction via excluded volume correction and Hagedorn exponentially rising density of states\cite{vovchenko}. Thus, as we pointed out in our previous work\cite{guruarxive}, the effects of Hagedorn density of states can be alternatively simulated by the EHRG with temperature and density dependent hadron masses. [Fig. \ref{thermo3}(b)] shows scaled entropy density in two models. We note that the entropy density rises more rapidly in IEHRG than in EHRG. This behavior of the entropy density plays crucial role in the behavior of the sound velocity as shown in Fig. (\ref{thermo5}) where we note that the sound velocity in IEHRG is smaller than that in EHRG. Sound wave is nothing but uniform compressions and expansions that propagate through medium with certain speed. In the compression or expansion, the thermodynamic system goes out of its equilibrium state and hence the internal processes set up in the system which try to restore the equilibrium. But the restoration processes which tries to bring the system towards equilibrium are not reversible whence accompanied by entropy production and energy dissipation. Greater the entropy production, less rapidly the local disturbances travel through the medium i.e smaller the speed of sound. Thus, more rapid rise of entropy density in IEHRG  makes speed of sound smaller than in EHRG.
 
\begin{figure}[h]
\vspace{-0.4cm}
\begin{center}
 \includegraphics[width=8cm,height=8cm]{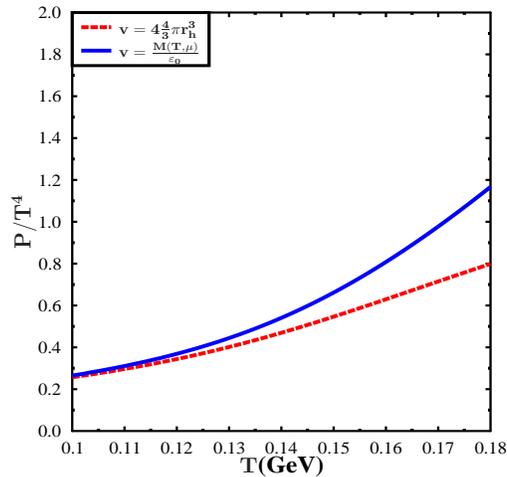}
  \caption{(Color online) Scaled pressure as function of temperature with mass dependent proper volume parameter (blue solid curve) and the constant proper volume parameter (red dashed curve). For the constant proper volume, we take uniform value for hard core radius $r_{h}=0.5$ fm for all the hadrons.} 
\label{thermo4}
  \end{center}
 \end{figure}

 Observation that all the thermodynamical quantities in IEHRG are numerically larger than in EHRG can be attributed to two factors. One is the Boltzmann factor Exp$(-M(T,\mu)/T)$ and other is the suppression factor $\Gamma^{-1}$. Boltzmann factor is a measure of probability that a given hadron is thermally excited at given T and $\mu$ and thus making contribution to the thermodynamical quantities at that T and $\mu$. In IEHRG model, masses of all but light mesons ($\pi$, K, $\eta$) decreases with temperature (and chemical potential)\cite{guruarxive}. Thus hadrons are excited more abundantly and easily in IEHRG than in EHRG where the hadron masses are constant independent of T and $\mu$. Besides, at given T and $\mu$, the suppression factor is always less than one and keep on decreasing with temperature because as the temperature increases, more and more hadrons are thermally excited occupying more and more system volume due to their finite size. Contrary to EHRG case, although the hadrons are produced more abundantly with T in IHRG, their hardcore radius is also depends on T and in fact it decreases with T due to our mass dependent parametrization of the proper volume which in turn depends on T and $\mu$. This effect can be noted from Fig. (\ref{thermo4}) which shows the scaled pressure in EHRG with T and $\mu$ dependent hadron masses and the mass dependent proper volume (IEHRG) as well as EHRG with T and $\mu$ dependent hadron mases but with the constant proper volume. We note that the scaled pressure is smaller with constant proper volume than that of IEHRG. This is the reflection the fact that the suppression factor decreases more leisurely in IEHRG than that in EHRG.

 Fig. (\ref{thermo5}) shows behavior of speed of sound in EHRG and IEHRG. We note that the sound velocity computed within IEHRG agrees with the lattice quantum chromoynamics\cite{prasad} quit well over wide range of temperatures. We further note that although the general behavior of C${_s}^{2}$ is same in two models at low temperatures, it differs quit significantly at high temperatures. In EHRG sound velocity rises very rapidly while it flattens out in IEHRG at high temperatures. As it has been pointed out in Ref. \cite{prakashvenu}, such large and steady rise in sound velocity is sufficient to indicate the acausal behavior typical of all excluded volume models. This acausal behavior of sound speed in excluded volume is not difficult to understand. Physically, the speed of sound is a measure of the efficiency of the medium to propagate the small disturbances as a longitudinal wave. In excluded volume models at low temperatures and low density where the system is dominated by light mesons, gas of hadrons can be treated as an compressible fluid which renders small and finite value of the speed of sound.  While at high temperatures and baryon densities, huge number of hadrons are thermally excited which tends to occupy the system volume more due to their finite size. Thus, at high temperatures and densities the gas of hadrons approach towards its incompressible liquid phase where the compressibility of the gas of hadrons approaches close to zero due to their close packing. Since speed of sound is inversely proportional to the compressibility, it start to diverges and even exceeds the speed of light whence violating causality. In contrast, although the hadrons are abundantly excited thermally in IEHRG model, their hardcore radius decreases with temperature due to our mass dependent parametrization of the proper volume. Whence, at high temperatures the gas of hadrons still remains compressible thus avoiding liquid-gas phase transition which renders small and finite value of sound speed. 

\section{Summary and Conclusion}
In present work, we suggested the possibility of curing the acausal behavior of the sound velocity in excluded volume HRG models by improving EHRG models which accounts for temperature and density dependent hadrons masses. We use Nambu-Jona-Lasinio model to obtain T and $\mu$ dependence of constituent quarks and light mesons. For heavier hadrons we use linear scaling rule in terms of constituent quarks. All the thermodynamical quantities in IEHRG are numerically larger than that in EHRG. This observation is attributed to two factors, $viz.$, Boltzmann factor Exp$-(M(T,\mu)/T)$ and suppression factor $\Gamma^{-1}$. Due to temperature and density dependent hadrons masses which actually decreases with these external parameters, hadrons are abundantly and easily excited at high temperature while simultaneously the suppression factor decreases less rapidly from one than that in EHRG model. By and by, all these effects are reflected in sound velocity where we found that the it flattens out at high temperatures in improved excluded volume HRG (EIHRG) whereas it rises very rapidly in normal excluded volume HRG(EHRG). The flattening of sound velocity may be attributed to the fact that the proper volume parameter in our model decreases as temperature  and density increases whence avoiding the possible liquid-gas phase transition.

 \newpage
 \def\karschkharzeev{F. Karsch, D. Kharzeev, and K. Tuchin, Phys. Lett. B
{\bf 663}, 217 (2008).}
\def\joglekar{J.C. Collins, A. Duncan, S.D. Joglekar, Phys. Rev. D {\bf 16}, 
438 (1977).}
\def\blaschke{J. Jankowski, D. Blaschke, M.Spalinski, Phys.Rev.D {\bf 87}, 105018
(2013). }
\def\gorenstein{M. Gorenstein, M. Hauer, O. Moroz, Phys.Rev.C {\bf 77}, 024911 (2008)}
\def\bugaev{K. Bugaev et al, Eur.Phys.J. A {\bf 49}, 30 (2013)}
\def\cpsingh{S.K. Tiwari, P.K. Srivastava, C.P. Singh, Phys.Rev. C {\bf 85},
014908 (2012).}
\def\chen{J.-W. Chen, Y-H. Li, Y.-F. Liu, and E. Nakano, Phys. Rev. D {\bf 76},
114011 (2007)}
\def\chennakano{J.-W. Chen, and E. Nakano, Phys. Lett. B {\bf 647}, 371 (2007)}
\def\itakura{K. Itakura, O. Morimatsu, and H. Otomo, Phys. Rev. D {\bf 77}, 014014
(2008)}
\def\cleymans{J. Cleymans, H. Oeschler, K. Redlich, and S. Wheaton, Phys.
Rev. C {\bf 73}, 034905 (2006)}
\def\hirano{P. Romatschke and U. Romatschke, Phys. Rev. Lett. {\bf 99}, 172301 (2007); T. Hirano and M. Gyulassy, Nucl. Phys. {\bf A 769}, 71, (2006).} 
\def\kss{P. Kovtun, D.T. Son and A.O. Starinets, Phys. Rev. Lett. {\bf 94},
 111601 (2005).}
\def\sasakiqp{C. Sasaki and K.Redlich,{\PRC{79}{055207}{2009}}.}
\def\sasakinjl{C. Sasaki and K.Redlich,{\NPA{832}{62}{2010}}.}
\def\ellislet{I.A. Shushpanov, J. Kapusta and P.J. Ellis,{\PRC{59}{2931}{1999}}
; P.J. Ellis, J.I. Kapusta, H.-B. Tang,{\PLB{443}{63}{1998}}.}
\def\prakashwiranata{A. Wiranata and Madappa Prakash, Phys. Rev. C {\bf 85}, 054908 (2012).}
\def\purnendu{P. Chakraborty and J.I. Kapusta {\PRC{83}{014906}{2011}}.}
\def\greco{S.Plumari,A. Paglisi,F. Scardina and V. Greco,{\PRC{83}{054902}{2012}.}}
\def\bes{H. Caines, arXiv:0906.0305 [nucl-ex], 2009.}
\def\greinerprl{J. Noronha-Hostler,J. Noronha and C. Greiner, 
{\PRL{103}{172302}{2009}}.}
\def\greinerprc{J. Noronha-Hostler,J. Noronha and C. Greiner
, {\PRC{86}{024913}{2012}}.}
\def\igorgreiner{J. Noronha-Hostler, C. Greiner and I. Shovkovy,
, {\PRL{100}{252301}{2008}}.}
\def\nakano{J.W. Chen,Y.H. Li, Y.F. Liu and E. Nakano,
 {\PRD{76}{114011}{2007}}.}
\def\itakura{K. Itakura, O. Morimatsu, H. Otomo, {\PRD{77}{014014}{2008}}.}
\def\wang{M.Wang,Y. Jiang, B. Wang, W. Sun and H. Zong, Mod. Phys. lett.
{\bf A76}, 1797,(2011).}
\def\rischkegorenstein{.D.H. Rischke, M.I. Gorenstein, H. Stoecker and
W. Greiner, Z.Phys. C {\bf 51}, 485 (1991).}
\def\hmnjl{Amruta Mishra and Hiranmaya Mishra, {\PRD{74}{054024}{2006}}.}
\def\pdgb{C. Amseler {\it et al}, {\PLB{667}{1}{2008}}.}
\def\shuryak{E.V. Shuryak, Yad. Fiz. {\bf 16},395, (1972).}
\def\leupold{S. Leupold, J. Phys. G{\bf32},2199,(2006)}
\def\peter{A. Andronic, P. Braun-Munzinger , J. Stachel and M. Winn,
{\PLB{718}{80}{2012}}}
\def\blum{M. Blum, B. Kamfer, R. Schluze, D. Seipt and U. Heinz,{\PRC{76}{034901}{2007}}.}
\def\jaminplb{M. Jamin{\PLB{538}{71}{2002}}.}
\def\ghosh{Sabyasachi Ghosh{\PRC{90}{025202}{2014}}.}
\def\csernai{L.P. Csernai, J.I. Kapusta and L.D. McLerran,{\PRL{97}{152303}{2006}}.}
\def\hagedorn{R. Hagedorn, Nuovo Cim. Suppl. 3,147 (1965); Nuovo Sim. A56,1027 (1968).}
\def\torieri{G. Torrieri and I. Mishustin,{\PRC{77}{034903}{2008}}.}
\def\fernandez{D. Fernandiz-Fraile and A.G. Nicola,{\PRL{102}{121601}{2009}}.}
\def\caron{S.Caron,{\PRD{79}{125009}{2009}}.}
\def\latticemeyer{H.B. Meyer,{\PRL{100}{162001}{2008}}.} 
\def\romatschke{P.Romatscke and D.T. Son,{\PRD{80}{065021}{2009}}.}
\def\moore{G.D. Mooore and O. Sarem, J. High Energy Phys. JHEP0809(2008)015.}
\def\dobado{A.Dobado and J. M. Torres-Rincon {\PRD{86}{074021}{2012}}.}
\def\daniel{D. Fernandez-Fraile, {\PRD{83}{065001}{2011}}.}
\def\gurunpa{G.P. Kadam, H. Mishra, Nuclear Physics A {\bf 934}, 133 (2015).}
\def\gurumpla{G.P. Kadam, Mod.Phys.Lett. A {\bf 30},  1550031 (2015).}
\def\demir{N. Demir, S.A. Bass, Phys. Rev. Lett. {\bf 102}, (2009) 172302.}
\def\phsd{V. Ozvenchuk, O. Linnyk, M.I. Gorenstein, E.L. Bratkovskaya, W. Cassing, Phys. Rev. C {\bf 87}, (2013) 064903.}
\def\ehrgrishke{D. H. Rischke, M. I. Gorenstein, H. Sto ̈cker, and W. Greiner,
Z. Phys. C {\bf 51}, 485 (1991).͒}
\def\ehrgclaymans{J. Cleymans, M. I. Gorenstein, J. Stalnacke, and E. Suhonen,
Phys. Scr. {\bf 48}, 277 (1993͒).}
\def\gavin{S. Gavin,  Nucl.Phys. A {\bf 435}, 826 (1985).}
\def\cannoni{Mirco Cannoni, Phys. Rev. D {\bf 89}, 103533 (2014).}
\def\gelmini{P. Gondolo and G. Gelmini, Nucl. Phys. B {\bf 360}, 145 (1991).}
\def\jeon{G.S. Denicol, C. Gale,  S. Jeon,  and J. Noronha, Phys. Rev. C {\bf 88}, 064901 (2013).}
\def\sinha{A. Buchel, R. C. Myers and A. Sinha, JHEP {\bf 0903}, 084 (2009).}
\def\heppe{P. Braun-Munzinger, I. Heppe, J. Stachel, Phys. Lett. B {\bf 465}, 15 (1999).}
\def\bugaev{K.A. Bugaev, D.R. Oliinychenko, A.S. Sorin, and G.M. Zinovjev, arxive:1208.5968v1.}
\def\cleymans{J. Cleymans, H. Oeschler, K. Redlich, S. Wheaton, Phys. Rev. C {\bf 73}, 034905 (2006).}
\def\moore{E. Lu, G. D. Moore, Phys. Rev. C {\bf 83}, 044901 (2011).}
\def\amseler{C. Amseler, et al., Phys. Lett. B {\bf 667}, 1 (2008).}
\def\moroz{O. Moroz, Ukr.J.Phys. {\bf 58}, 1127 (2013).}
\def\yen{G. D. Yen, M. I. Gorenstein, W. Greiner, and S. N. Yang, Phys.Rev. C {\bf 56}, 2210 (1997).}
\def\sarkarghosh{ S. Ghosh, G. Krein, S. Sarkar, Phys.Rev. C {\bf 89}, 045201 (2014).}
\def\finazzo{ S. I. Finazzo, R. Rougemont, H. Marrochio, J. Noronha, JHEP {\bf 1502}, 051 (2015).}
\def\khvorostukhin{A.S. Khvorostukhin, V.D. Toneev, D.N. Voskresensky, Nucl. Phys. A {\bf 845}, 106 (2010).}
\def\sghosh{S. Ghosh, Int. J. Mod. Phys. A {\bf 29}, 1450054 (2014).}
\def\sghoshnucl{S. Ghosh, Phys. Rev. C {\bf 90}, 025202 (2014).}
\def\sghoshnjl{S. Ghosh,  A. Lahiri, S. Majumder, R. Ray, S. K. Ghosh, Phys. Rev. C {\bf 88}, 068201 (2013).}
\def\cohen{ T.D. Cohen, Phys. Rev. Lett. {\bf 99}, 021602 (2007).}
\def\rebhan{ A. Rebhan and D. Steineder, Phys. Rev. Lett. {\bf 108}, 021601 (2012).}
\def\mamo{ K.A. Mamo, JHEP {\bf 70}, 1210 (2012).}
\def\weise{R. Lang, N. Kaiser, and W. Weise, arxive:1506.02459}
\def\dobadojuan{A. Dobado, F. J. Llanes-Estrada and J. M. Torres-Rincon, Phys. Lett. B {\bf 702}, 43 (2011).}
\def\gyulassy{P. Danielewicz and M. Gyulassy, Phys.Rev. D {\bf 31}, 53 (1985).}
\def\borsanyi{S. Borsányi, et al., JHEP {\bf 1011}, 077 (2010).}
\def\prasad{Bazavov et al., Phys. Rev. D {\bf 90}, 094503 (2014).}
\def\WPblk{A. Wiranata and M. Prakash, Nucl. Phys. A {\bf 830}, 219–222 (2009).}
\def\prakash{M. Prakash, M. Prakash, R. Venugopalan and G. Welke, Phys.Rept. {\bf 227}, 321-366 (1993).}
\def\wiranatakoch{A. Wiranata, V. Koch and  M. Prakash, X.N. Wang, J.Phys.Conf.Ser. {\bf 509}, 012049 (2014).}
\def\wiranataprc{A. Wiranata and M. Prakash, Phys.Rev. C {\bf 85}, 054908 (2012).}
\def\wiraprapurn{ A. Wiranata, M. Prakash and  P. Chakraborty, Central Eur.J.Phys. {\bf 10}, 1349-1351 (2012).}
\def\wiranatademir{N. Demir and A. Wiranata, J.Phys.Conf.Ser. {\bf 535}, 012018 (2014).}
\def\gorensteinplb{M. Gorenstein, V. Petrov and G. Zinovjev, Phys.Lett. B {\bf 106}, 327-330 (1981).}
\def\goregranddonprc{G. D. Yen, M. Gorenstein, W. Greiner and Shin Nan Yang, Phys. Rev. C {\bf 56}, 2210 (1997).}
\def\gorensteinjpg{M. Gorenstein, H. St\"{o}ker, G. D. Yen, Shin Nan Yang and W. Greiner, J. Phys. G {\bf 24}, 1777 (1998).}
\def\yengorenprc{G.D. Yen and M. Gorenstein, Phys. Rev. C {\bf 59}, 2788 (2009).}
\def\gorengreinerjpg{M. Gorenstein, W. Greiner and Shin Nan Yang, J. Phys. G {\bf 24}, 725 (1998).}
\def\gorengreinerprc{M. Gorenstein, M. Ga'zdicki and W. Greiner, Phys. Rev. C {\bf 72}, 024909 (2005).}
\def\tawfik{A. Tawfik and M. Wahba, Ann. Phys. {\bf 522}, 849-856 (2010).}
\def\jeonyaffe{S. Jeon and L. Yaffe, Phys.Rev. D {\bf 53}, 5799-5809 (1996).}
\def\liao{J. Liao, V. Koch, Phys. Rev. D {\bf 81}, 014902 (2010).}
\def\Khov{A. Khvorostukhin, V. Toneev, D. Voskresensky, Nucl. Phys. A {\bf 845}, 106–146, (2010).}
\def\nicola{D. Fernandez-Fraile, A. Gomez-Nicola, Eur.Phys.J. C {\bf 62}, 37-54 (2009).}
\def\Huovinen{P. Huovinen, P. Petrecky, Nucl. Phys. A {\bf 837}, 26 (2010).}
\def\borsonyi{S. Borsányi, et al., JHEP {\bf 1009}, 073 (2010).}
\def\GMOR{J. Gasser and H. Leutwyler, Nucl. Phys. B{bf\ 250}, 465 (1985).}
\def\Huovinen{P. Huovinen, P. Petrecky, Nucl. Phys. A {\bf 837}, 26 (2010).}
\def\welke{G. M. Welke, R. Venugopalan, M. Prakash, Phys. Lett. B {\bf 245}, 137 (1990).}
\def\prakashvenu{R. Venugopalan, M. Prakash, Nucl. Phys. A {\bf 546}, 718 (1992).}
\def\andronic{A. Andronic, P. Braun-Munzinger, J. Stachel, M. Winn, Phys. Lett. B {\bf 718}, 80 (2012).}
\def\kapusta{M. Albright, J. Kapusta, and C. Young, Phys. Rev. C {\bf 90}, 024915 (2014).}
\def\bhattacharyya{Abhijit Bhattacharyya, Rajarshi Ray, Subhasis Samanta, and Subrata Sur, Phys. Rev. C {\bf 91}, 041901 (2015).}
\def\garg{ P. Garg,  D.K. Mishra, P.K. Netrakanti, B. Mohanty, A.K. Mohanty, B.K. Singh, N. Xu, Phys.Lett. B {\bf 726}, 691 (2013).}
\def\guruprc{G.P. Kadam, H. Mishra, Phys. Rev.C {\bf 92}, 035203 (2015).}
\def\costa{P. Costa, M.C. Ruivo, C.A. de Sousa, Phys. Rev. C {\bf 70}, 025204 (2004).}
\def\klevansky{S. P. Klevansky, Rev. Mod. Phys. {\bf 64}, 649 (1992).}
\def\hatsuda{T. Hatsuda and T. Kunihiro, Phys. Rep. {\bf 247}, 221 (1994).}
\def\vogl{U. Vogl and W. Weise, Prog. Part. Nucl. Phys. {\bf 27}, 195 (1991).}
\def\hufner{P. Rehberg, S. P. Klevansky, and J. H\"{u}fner, Phys. Rev. C {\bf 53}, 410 (1996).}
\def\leupold{S. Leupold, J. Phys. G {\bf 32}, 2199 (2006).}
\def\jankowski{J. Jankowski, D. Blaschke, and M. Spali\'{n}ski, Phys. Rev. D {\bf 87}, 105018 (2013).}
\def\vovchenko{V. Vovchenko, D. V. Anchishkin and M. I. Gorenstein, Phys. Rev. C {\bf 91}, 024905 (2015)}
\def\fodor{S. Borsanyi, Z. Fodor, C. Hoelbling, S. D. Katz, S. Krieg, and K. K. Szabo, Phys. Lett. B {\bf 730}, 99 (2014).}
\def\cleymansatz{J. Cleymans and H. Satz, Z. Phys. C {\bf 57}, 135 (1993).}
\def\magestro{P. Braun-Munzinger, D. Magestro, K. Redlich, and J. Stachel,
Phys. Lett. B {\bf 518}, 41 (2001).}
\def\rafelski{J. Rafelski and J. Letessier, Nucl. Phys. A {\bf 715}, 98 (2003).}
\def\stachelandronic{A. Andronic, P. Braun-Munzinger, and J. Stachel, Nucl. Phys.
A 772, 167 (2006).}
\def\Becattini{F. Becattini, J. Cleymans, A. Keranen, E. Suhonen, and
K. Redlich, Phys. Rev. C 64, 024901 (2001).}
\def\torres{J.M. Torres-Rincon, B. Sintes and J. Aichelin, Phys. Rev. C {\bf 91}, 065206 (2015).}
\def\DBM{R. Dashen, S.-K. Ma, and H. J. Bernstein, Phys. Rev. {\bf 187}, 349 (1969).}
\def\kapustaolive{J. Kapusta, K. Olive, Nucl. Phys. A {\bf 408}, 478 (1983).}
\def\Bugaev{K. A. Bugaev, Nucl. Phys. A {\bf807}, 251 (2008).}
\def\guruarxive{G.P. Kadam, H. Mishra, arXiv:1509.06998 [hep-ph].}

\end{document}